\documentclass[twocolumn,aps,prl,showpacs,superscriptaddress]{revtex4}
\usepackage{graphicx}
\usepackage{amsfonts}
\usepackage{amsmath}
\usepackage{amssymb}

\def\endproof{\vrule height6pt width6pt depth0pt}

\begin{document}

\title{State-independent quantum contextuality and maximum nonlocality}


\author{Ad\'an Cabello}
 \affiliation{Departamento de F\'{\i}sica Aplicada II, Universidad de
 Sevilla, E-41012 Sevilla, Spain}
 \affiliation{Department of Physics, Stockholm University, S-10691
 Stockholm, Sweden}


\date{\today}



\begin{abstract}
Recently, Yu and Oh [Phys. Rev. Lett. \textbf{108}, 030402 (2012)] have conjectured that the simplest set of vectors needed to prove state-independent contextuality on a qutrit requires $13$ vectors. Here we first prove that a necessary and sufficient condition for a set of vectors in any dimension $d \ge 3$ to prove contextuality for a system prepared in a maximally mixed state is that the graph $G_C$ in which vertices represent vectors and edges link orthogonal ones has chromatic number $\chi(G_G)$ larger than $d$. Then, we prove Yu and Oh's conjecture. Finally, we prove that any set satisfying $\chi(G_G)>d$ assisted with $d$-party maximum entanglement generates nonlocality which cannot be improved without violating the no-signalling principle. This shows that any set satisfying $\chi(G_G)>d$ is a valuable resource for quantum information processing.
\end{abstract}



\maketitle


One of the most intriguing aspects of the quantum world is the so-called state-independent contextuality (SIC) of results. For every physical system of dimension $d \ge 3$ obeying the laws of quantum mechanics (QM), there is always a finite set of tests whose results cannot be assigned independently of which other tests are performed. Unlike quantum nonlocality, which can be proven only for composite systems in entangled states, quantum contextuality may be state-independent: For any $d \ge 3$, the same set of tests proves contextuality for any state of the system.

For years, SIC has been based on a mathematical result pointed out by Kochen and Specker (KS) \cite{Specker60,KS67} and Bell \cite{Bell66}: the existence of sets of projectors onto vectors in dimension $d \ge 3$, representing yes-no tests, for which there is no map $f:\mathbb{C}^d \rightarrow \{1,0\}$, i.e., no assignment of results yes or no, such that for all orthonormal bases $b \subseteq \mathbb{C}^d$, $\sum_{v \in b} f(v)=1$. These sets are called KS sets. The first KS set had 117 vectors in $\mathbb{C}^3$ \cite{KS67}; the simplest known KS set in $\mathbb{C}^3$ has 31 vectors \cite{Peres95}, and the simplest KS set has 18 vectors (in $\mathbb{C}^4$) \cite{CEG96}. Compact proofs of SIC based on Pauli operators rather than vectors, such as those introduced by Peres and Mermin \cite{Peres90,Mermin90}, can always be expressed in terms of KS sets: The common eigenstates of the commuting Pauli operators constitute a KS set of vectors \cite{Peres91,KP95}. This fundamental role attributed to KS sets explains the interest in methods to obtain new KS sets \cite{ZP93,CG96,CEG05,PMMM05}, games with quantum advantage based on KS sets \cite{HR83,RW04}, experimental tests of SIC based on KS sets \cite{Cabello08,KZGKGCBR09,BBCP09,ARBC09,MRCL09}, and Bell inequalities based on KS sets \cite{Cabello01,YZZYZZCP05,CBPMD05,AGACVMC11}.

For all these reasons, Yu and Oh's recent observation \cite{YO11} that a set of vectors which is {\em not} a KS set can be used to prove SIC came up as a surprise. Yu and Oh show that a noncontextuality inequality (i.e., a correlation inequality satisfied by any noncontextual theory) using 13 vectors in $\mathbb{C}^3$ for which the KS map {\em does exist} works equally well as a SIC proof. Moreover, Yu and Oh conjectured that this 13-vector set is the one with the smallest number of vectors needed to prove SIC. Another example of a set of vectors which is not a KS set but proves SIC is the 21-vector set in Ref.~\cite{BBC11}.

Some natural questions arise as a result of these examples: What is the common property of all KS sets and the sets in Refs.~\cite{YO11} and \cite{BBC11}? What is necessary for a set of vectors to prove SIC? Which of these proofs is the simplest? What is the connection between these proofs and quantum nonlocality and its applications for information processing? The aim of this Letter is to answer all these questions.


{\bf Theorem 1 (Necessary condition for SIC): }{\em A necessary condition for a set $C=\{v_i\} \subset \mathbb{C}^d$ (representing a set of yes-no tests) to reveal SIC is that the graph $G_C$ in which the vertices are the vectors in $C$ and edges link vectors if and only if they are orthogonal has chromatic number greater than $d$.}

The chromatic number of a graph $G$, denoted by $\chi(G)$, is the minimum number of colors $d$ such that there exists a proper $d$-coloring of $G$. A proper $d$-coloring of a graph is an assignment of $d$ colors to the vertices of the graph such that adjacent vertices have different colors.


{\em Proof: }Any SIC proof must identify a contradiction between noncontextual hidden variable theories (NCHVTs) and the predictions of QM for any possible state; in particular, for a maximally mixed state. If the predictions refer to a maximally mixed state and only involve the vectors in $C$, then they must be expressed as some restriction to the possible colorings of the vertices of $G_C$. We will show that, if $\chi(G_C) = d$, then there are colorings which do not contradict any prediction of QM for this state. However, if $\chi(G_C) > d$, then there is one prediction of QM which NCHVTs cannot reproduce.

Consider a $d$-dimensional system in the state $\rho=\frac{\openone}{d}$ obtained after tracing out $d-1$ particles of the $d$-particle $d$-level system prepared in the supersinglet state \cite{Cabello03}
\begin{equation}
|{\cal S}_d^{(d)} \rangle={1 \over \sqrt {d!}}
\sum_{\scriptscriptstyle{ {\stackrel{\scriptscriptstyle{\rm
permutations}} {{\rm of}\;01\ldots(d-1)}}}} \!\!\!\!\!\!
(-1)^t|ij\ldots d \rangle,
\label{SNSN}
\end{equation}
where $t$ is the number of transpositions of pairs of elements that must be composed to place the elements in canonical order (i.e., $0,1,2,\ldots,d-1$). For example, if $d=3$,
\begin{equation}
|{\cal S}_3^{(3)} \rangle = {1 \over \sqrt {6}} (
| 012 \rangle - | 021 \rangle -
| 102 \rangle + | 120 \rangle + | 201 \rangle - | 210 \rangle ).
\end{equation}
${\cal S}_d^{(d)}$ is the state of total spin zero of $d$ particles of spin $(d-1)/2$. For us, the important property of this state is that it is $d$-lateral unitary invariant, i.e., the tensor product of $d$ equal unitary operators $U \in SU(d)$ acting on ${\cal S}_d^{(d)}$ produces the same state (within a possible phase factor which is physically irrelevant). That is, for any $U \in SU(d)$, $U^{\bigotimes d} |{\cal S}_d^{(d)}\rangle=|{\cal S}_d^{(d)}\rangle$.

Therefore, if the $d$ particles are distributed among $d$ distant parties, and each of them measures the $d$-outcome observable
\begin{equation}
 O^{\{\lambda_i\}} = \sum_{v_i\in b} \lambda_i |v_i\rangle \langle v_i|,
 \label{observables}
\end{equation}
where $b$ is an orthogonal basis in $C \cup C'$, where $C'$ is the set of vectors not included in $C$ but needed to complete orthogonal bases using two ore more vectors of $C$. Then, each party will obtain a different result $\lambda_i$, and their reduced state will collapse to a different $v_i \in C \cup C'$. The same holds {\em for every observable} $O^{\{\lambda_i\}}$, since $|{\cal S}_d^{(d)} \rangle$ has an identical expression in any basis. This property implies that, any NCHVT reproducing QM should assign each $v_i \in C$ to only one party, satisfying that orthogonal $v_i$'s should be assigned to different parties. However, this assignment is impossible if $\chi(G_C) > d$, since it would constitute a proper $d$-coloring of $G_C$.

Reciprocally, if a proper $d$-coloring exists, then it is possible to assign each $v_i \in C$ to one party, satisfying that orthogonal $v_i$'s are assigned to different parties; thus NCHVTs exist which do not contradict {\em any} of the predictions of QM for the state $\rho=\frac{\openone}{d}$ and the observables ($|v_i\rangle \langle v_i|$ and $O^{\{\lambda_i\}}$) defined from $C$.\hfill \endproof


Graphs with $\chi(G_C) > d$ do not exist in $d=2$ (where the only possible connected graph is the one with two linked vertices), but exist in $d=3$ and higher dimensions.

Every KS set has $\chi(G_C)>d$. To prove it, notice that a KS set is one in which one cannot assign two colors, red and green, such that: (i) for every $d$ mutually linked vertices, one of them is green and the other are all red, and (ii) two greens cannot be adjacent. Then, it is also impossible to assign $d$ colors, green, red$_1$,\ldots, red$_{d-1}$, satisfying (i), (ii), and the extra constraint that two red$_i$'s cannot be adjacent.

However, the converse is not true: There are sets with $\chi(G_C)>d$ which are not KS sets. For instance, the sets in Refs.~\cite{YO11} and \cite{BBC11} are not KS sets, but both have $\chi(G_C)=4$ and $d=3$.


{\bf Theorem 2 (Yu and Oh's conjecture \cite{YO11}): }{\em The smallest set of vectors in $\mathbb{C}^3$ needed to prove SIC is the 13-vector set $S_3$ containing $(0,0,1)$, $(0,1,1)$, $(0,1,-1)$, $(1,1,1)$, $(1,1,-1)$, and all the vectors obtained by permuting their components.}

{\em Proof: }Let us denote by $\xi(G)$ the minimum $d$ such that there exists a orthogonal representation of $G$ in $\mathbb{C}^d$. An orthogonal representation of a graph in $\mathbb{C}^d$ is an assignment of vectors in $\mathbb{C}^d$ to the vertices of the graph such that two vertices are adjacent if and only if the vectors are orthogonal. There is no graph $G$ with less than 13 vertices such that $\xi(G) \leq 3 < \chi(G)$. This can be checked by generating all non-isomorphic connected square-free graphs (since a square cannot represent the orthogonality relations of four different rays in $\mathbb{C}^3$), and then selecting those such that $\chi(G)>3$. There is only one graph with less than 13 vertices such that $\chi(G)>3$, but it contains a 10-vertex subgraph which cannot represent $10$ different rays in $\mathbb{C}^3$ (see Appendix).\hfill \endproof


$S_3$ is a subset of the 33-vector KS set in Ref.~\cite{Peres91} and its special noncoloring properties were pointed out in Ref.~\cite{Cabello96}. The fact that the graph $G_C$ associated to $C=S_3$ has $\xi(G_C) < \chi(G_C)$ was observed in Ref.~\cite{Prosolupov04}. A conjecture about $S_3$ similar to Yu and Oh's can be found in an unpublished paper by DeVos {\em et al.} \cite{DGGMN09}. $S_3$ has also attracted attention in relation to the rank-1 quantum chromatic number \cite{SS11}.

The condition $\chi(G_C) > d$ is necessary and sufficient for proving contextuality, assuming that the system is in a maximally mixed state. However, it is not sufficient for SIC when one of the vectors in $C$ is orthogonal to all the other vectors in $C$. For example, consider the set $S_{3+j}=\{(x,0,\ldots,0) \in \mathbb{C}^{3+j}\;|\;x \in S_3\} \cup \{(0,0,0,1,0,\ldots,0)$, $(0,0,0,0,1,0,\ldots,0),\ldots$, $(0,\ldots,0,1) \in \mathbb{C}^{3+j}\}$, with $j=0,1,2\ldots$ The dimension is $d=3+j$ and the corresponding graph has chromatic number $4+j$. However, if the system is prepared in the state $(0,0,0,1,0,\ldots,0)$, then there is a NCHVT which fulfils all the predictions of QM.

Interestingly, the necessary and sufficient condition for contextuality of the maximally mixed state plays a fundamental role in quantum nonlocality.


{\bf Theorem 3 ($\chi(G_C)>d$ and nonlocal games): }{\em Every set $C \subset \mathbb{C}^d$ with graph $G_C$ such that $\chi(G_C)>d$ can be converted to a nonlocal game.}

{\em Proof: }The following game provides an example: $d$ mutually incommunicated players claim that they can assign each of the $v_i \in C$ to one of the players, satisfying the condition that orthogonal vectors are always assigned to different parties. To test this claim, in each run a referee chooses a basis in $C \cup C'$ for each player and asks him which vector of this basis he has. The players win if two conditions are simultaneously achieved. (i) When the referee chooses two bases $b=\{v_i,v_j,\ldots,v_d\}$ and $b'=\{v_i,v'_j,\ldots,v'_d\}$, one for each player, with a common vector $v_i$, then these two players must never answer both $v_i$. In particular, this implies that, when the referee chooses the same basis $b$ for all players, all of them must answer a different vector of $b$. (ii) When the referee chooses the same basis $b$ for all players except one (and each of them answers a different vector of $b$), and chooses a basis $b'=\{v_i,v'_j,\ldots,v'_d\}$ which contains the vector $v_i$ of $b$ that none of the players has answered for the remaining player, then this player must always answer $v_i$.

No classical strategy allows the players to win this game in every instance, since $\chi(G_C) > d$ implies that a proper $d$-coloring of $G_C$ does not exist. However, the players can always win the game if, in each run, each of them has a $d$-level particle belonging to a $d$-particle system prepared in the supersinglet state $|{\cal S}_d^{(d)}\rangle$ given by \eqref{SNSN}, measures on his particle the $O^{\{\lambda_i\}}$ corresponding to the basis the referee is asking, and provides as answer the $v_i$ corresponding to the obtained result $\lambda_i$.\hfill \endproof

However, the quantum strategy for winning the game in every instance vanishes if the prepared quantum state or the local measurements are imperfect. Testing nonlocality in realistic laboratory conditions requires a Bell inequality.


{\bf Theorem 4 ($\chi(G_C)>d$ and Bell inequalities): }{\em Every set $C \subset \mathbb{C}^d$ with graph $G_C$ such that $\chi(G_C)>d$ can be converted into a Bell inequality violated by QM.}

{\em Proof: }Consider $d\ge3$ separated parties, each of them having a particle on which each party measures at random one of the $d$-outcome observables $O^{\{\lambda_i\}}$ defined in \eqref{observables}. The result of any of these measurements can be labeled as $\lambda_i$, with $i=1,\ldots$, card$(C \cup C')$. If all measurements are spacelike separated, then the following Bell inequality holds:
\begin{widetext}
\begin{equation}
\begin{split}
 &\sum_{O^{\{\lambda_i,\lambda_j,\ldots,\lambda_d\}},O^{\{\lambda_i,\lambda_j',\ldots,\lambda_d'\}}} \sum_{p,q} \left[1-P\left(\lambda_i,\lambda_i|O_p^{\{\lambda_i,\lambda_j,\ldots,\lambda_d\}},O_q^{\{\lambda_i,\lambda_j',\ldots,\lambda_d'\}}\right)\right] \\
 &+\sum_{O^{\{\lambda_i,\lambda_j,\ldots,\lambda_d\}},O^{\{\lambda_i,\lambda_j',\ldots,\lambda_d'\}}} \sum_{p} \left[1-P\left(\neq \lambda_i,\lambda_j,\ldots,\lambda_d|O_p^{\{\lambda_i,\lambda_j',\ldots,\lambda_d'\}},O_q^{\{\lambda_i,\lambda_j,\ldots,\lambda_d\}},\ldots,O_d^{\{\lambda_i,\lambda_j,\ldots,\lambda_d\}}\right)\right] \stackrel{\mbox{\tiny{ LHV}}}{\leq} \Omega_{\rm{LHV}} \stackrel{\mbox{\tiny{ QM}}}{<} \Omega_{\rm{QM}},
 \label{Bellinequality}
 \end{split}
\end{equation}
\end{widetext}
where $P\left(\lambda_i,\lambda_i|O_p^{\{\lambda_i,\lambda_j,\ldots,\lambda_d\}},O_q^{\{\lambda_i,\lambda_j',\ldots,\lambda_d'\}}\right)$ is the joint probability that party $p$ measures $O^{\{\lambda_i,\lambda_j,\ldots,\lambda_d\}}$ and obtains $\lambda_i$, and party $q$ measures $O^{\{\lambda_i,\lambda_j',\ldots,\lambda_d'\}}$ and also obtains $\lambda_i$ [this probability must be zero according to rule (i) in the proof of Theorem 3], and the corresponding sums extend to all possible pairs of observables with a common eigenvector and all possible pairs $(p,q)$ of parties; the second probability, $P\left(\neq \lambda_i,\ldots|\ldots,O_d^{\{\lambda_i,\lambda_j,\ldots,\lambda_d\}}\right)$, is the joint probability that party $p$ measures $O^{\{\lambda_i,\lambda_j',\ldots,\lambda_d'\}}$ and finds a result different than $\lambda_i$, and all the other parties measure $O^{\{\lambda_i,\lambda_j,\ldots,\lambda_d\}}$ and obtain different results, all of them different than $\lambda_i$ [this probability must be zero according to rule (ii) in the proof of Theorem 3], and the corresponding sums extend to all to all possible pairs of observables with a common eigenvector and all parties $p$.

In QM, if the system is a $d$-level system in the state $|{\cal S}_d^{(d)}\rangle$ given by \eqref{SNSN}, and the parties measure the observables $O^{\{\lambda_i\}}$ given by \eqref{observables}, then all the joint probabilities in \eqref{Bellinequality} are zero, thus the left-hand side of \eqref{Bellinequality} attains its algebraic maximum given by the total number of joint probabilities appearing in \eqref{Bellinequality}. This is the maximum of the left-hand side of \eqref{Bellinequality} in QM, denoted by $\Omega_{\rm{QM}}$.

However, the maximum for local hidden variable theories, denoted by $\Omega_{\rm{LHV}}$ is strictly smaller than $\Omega_{\rm{QM}}$. This can be proven by noting that $\Omega_{\rm{LHV}}$ is always reached by some deterministic model in which a deterministic outcome is assigned to every local measurement, so all the probabilities in \eqref{Bellinequality} can thus only equal to 0 or 1. Suppose that a deterministic model attains $\Omega_{\rm{QM}}$, the model then specifies the outcomes for all local measurements such that every possible outcome is assigned to one party, and outcomes corresponding to orthogonal vectors are assigned to different parties. However, this is impossible, since this implies a proper $d$-coloring of $G_C$, which is impossible since we are assuming that $\chi(G_C)>d$. Therefore, one concludes that $\Omega_{\rm{LHV}} < \Omega_{\rm{QM}}$.\hfill \endproof

Hypothetical post-quantum theories not violating the no-signalling principle can give larger violations of the Clauser-Horne-Shimony-Holt Bell inequality than QM \cite{PR94}. However, there is a special class of Bell inequalities in which the maximum quantum violation is also the maximum violation allowed by the no-signalling principle. The corresponding quantum correlations are called ``fully nonlocal'' and have many applications for information processing (see \cite{AGACVMC11} and references therein). It has been recently pointed out that Bell inequalities showing maximum or full nonlocality can be constructed from KS sets \cite{AGACVMC11}. Here we prove a stronger result.


{\bf Theorem 5 ($\chi(G_C) > d$ and maximum nonlocality): }{\em Every set $C \subset \mathbb{C}^d$ with graph $G_C$ such that $\chi(G_C) > d$ can be used to reveal the maximum nonlocality allowed by the no-signalling principle.}

{\em Proof: }The maximum quantum violation of the Bell inequality in the proof of Theorem 4 satisfies the no-signalling principle and is also the maximum algebraic violation. Therefore, it is the maximum violation satisfying the no-signalling principle.\hfill \endproof


{\em Conclusions and open questions.---}In this Letter we have proven a recent conjecture by Yu and Oh \cite{YO11}, which allows us to conclude that the simplest proof of SIC in the simplest physical system capable of exhibiting SIC requires exactly 13 rank-1 projectors. This result will have impact in the design of experiments seeking SIC on qutrits and in the applications of SIC for information processing.

In addition, we have shown that there is a deeper reason, beyond the one pointed out by KS \cite{KS67} and Bell \cite{Bell66}, explaining why some sets of rank-1 projectors in QM reveal SIC. We have provided a necessary condition for a set of rank-1 projectors to be able to reveal SIC, and show that this condition is also sufficient to prove contextuality if the system is in the maximally mixed state. This condition is $\chi(G_C) > d$. We have also shown that sets of rank-1 projectors satisfying this condition exist in any $d \ge 3$, and provided explicit examples with $d+10$ projectors.

Moreover, we have shown that any set of projectors satisfying this condition, assisted with maximum $d$-particle $d$-level entanglement, permits a proof of nonlocality with one peculiarity: The nonlocality is maximum in the sense that it cannot be outperformed by any nonsignalling resources. This establishes a connection between the necessary and sufficient condition for contextuality for maximally mixed states and maximum nonlocality. A still open question is whether any proof of maximum nonlocality can always be associated to a set of projectors satisfying this condition.

These results take the connection between quantum contextuality and nonlocality a step further: It is known that contextuality together with entanglement provide nonlocality. Here we have shown that the existence of sets of observables capable to reveal contextuality for maximally mixed states is a resource which, together with maximum entanglement (a different resource), permits nonlocality that cannot be improved without violating the no-signalling principle.

Another interesting final question is the connection between sets with $\chi(G_C) > d$ and entanglement assisted zero-error capacity. It is known that KS sets can be used to increase the one-shot entanglement assisted zero-error capacity for classical messages \cite{CLMW10}. A reasonable conjecture is that the same should be possible for channels constructed from sets with $\chi(G_C) > d$.


\begin{acknowledgments}
The author thanks A. Ac\'{\i}n, E. Amselem, L. Aolita, M. Ara\'{u}jo, I. Bengtsson, K. Blanchfield, M. Bourennane, C. Budroni, R. Gallego, O. G{\"u}hne, M., P., and R. Horodecki, A. Kent, M. Kleinmann, J.-\AA. Larsson, A. J. L\'opez-Tarrida, G. Scarpa, and S. Severini for discussions, and J. R. Portillo for his help with {\tt nauty}. This work was supported by the Projects No.\ FIS2008-05596 and No.\ FIS2011-29400, and the Wenner-Gren Foundation.
\end{acknowledgments}


\section{Appendix}


\begin{figure}[t]
\centerline{\includegraphics[width=3.8cm]{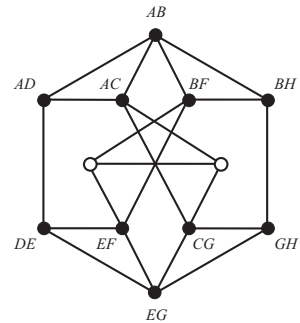}}
\caption{\label{Fig02}The only square-free graph with less than 13 vertices and $\chi(G)>3$. As shown in the Lemma, the 10-vertex subgraph of the black vertices cannot represent $10$ different rays in $\mathbb{C}^3$.}
\end{figure}


From all graphs with less than 13 vertices free of squares (which can be obtained using {\tt nauty} \cite{McKay90}), the only one such that $\chi(G)>3$ is shown in Fig. \ref{Fig02}. However, it contains a 10-vertex subgraph which cannot represent $10$ different rays in $\mathbb{C}^3$.

{\em Lemma: }It is impossible to have 10~different rays
$\hat{v}_{AB}$, $\hat{v}_{AC}$, $\hat{v}_{AD}$, $\hat{v}_{BF}$,
$\hat{v}_{BH}$, $\hat{v}_{CG}$, $\hat{v}_{DE}$, $\hat{v}_{EF}$,
$\hat{v}_{EG}$, and $\hat{v}_{GH}$ in $S^{2}$ such that rays
sharing one letter are orthogonal.


{\em Proof:} Without loss of generality, we can choose
$\hat{v}_{AB}=(1,0,0)$, $\hat{v}_{AC}=(0,1,0)$, and
$\hat{v}_{AD}=(0,0,1)$. Then, $\hat{v}_{BF}=(0,c_1,s_1)$,
$\hat{v}_{BH}=(0,s_1,-c_1)$, $\hat{v}_{CG}=(c_2,0,s_2)$,
$\hat{v}_{DE}=(c_3,s_3,0)$, where $c_i=\cos \theta_i$ and
$s_i=\sin \theta_i$. Then, $\hat{v}_{GH}=N_{GH}(-s_1 s_2,c_1
c_2,s_1 c_2)$ and $\hat{v}_{EF}=N_{EF}(s_1 s_3,-s_1 c_3,c_1 c_3)$,
where $N_{jk}$ are the corresponding normalization factors. If
$\hat{v}_{EG}$ must be orthogonal to $\hat{v}_{CG}$ and
$\hat{v}_{GH}$, then $\hat{v}_{EG}=N_{EG}(c_1 c_2 s_2,s_1,-c_1 c_2^2)$.
On the other hand, if $\hat{v}_{EG}$ must be orthogonal to
$\hat{v}_{DE}$ and $\hat{v}_{EF}$, then $\hat{v}_{EG}=N_{EG}(-c_1 c_3 s_3,c_1 c_3^2,s_1)$.
Therefore, $s_2^2=-s_3^2$, which only admits the
solution $s_2=s_3=0$, which is not admissible because $\hat{v}_{CG}$
and $\hat{v}_{AB}$ are assumed to be different rays.\hfill\endproof



\end{document}